\documentclass[12pt,preprint]{aastex} % one-column, double-spaced document:

\shorttitle{ Loop Temperatures}
\shortauthors{Schmelz et al.}

\begin{document}

\title{Are Coronal Loops Isothermal or Multithermal? Yes!}

\author{J.T. Schmelz, K. Nasraoui, L.A. Rightmire, J.A. Kimble, G. Del Zanna\altaffilmark{1}, J.W. Cirtain\altaffilmark{2}, E.E. DeLuca\altaffilmark{3}, H.E. Mason\altaffilmark{4}}
\affil{Physics Department, University of Memphis, Memphis, TN 38152} 
\email{jschmelz@memphis.edu}%; kaouther2001@yahoo.com}
\altaffiltext{1}{Mullard Space Science Laboratory,
University College London, Holmbury  St. Mary, Dorking, Surrey, RH5 6NT UK; G.Del-Zanna@damtp.cam.ac.uk}
\altaffiltext{2}{MSFC/NASA, NSSTC, 320 Sparkman Ave., Huntsville, AL 35805; Jonathan.W.Cirtain@nasa.gov}
\altaffiltext{3}{Harvard-Smithsonian Center for Astrophysics, 60 Garden St. Mail Stop 58, Cambridge, MA, 02138; edeluca@cfa.harvard.edu}
\altaffiltext{4}{DAMTP, Centre for Mathematical Sciences,
  Wilberforce Road, Cambridge CB3 OWA, UK; H.E.Mason@damtp.cam.ac.uk}

\begin {abstract}
Surprisingly few solar coronal loops have been observed simultaneously with TRACE and SOHO/CDS, 
and even fewer analyses of these loops have been conducted and published. The SOHO Joint 
Observing Program 146 was designed in part to provide the simultaneous observations required 
for in-depth temperature analysis of active region loops and determine whether these loops are 
isothermal or multithermal. The data analyzed in this paper were taken on 2003 January 17 of 
AR 10250. We used TRACE filter ratios, emission measure loci, and two methods of differential 
emission measure analysis to examine the temperature structure of three different loops. TRACE 
and CDS observations agree that Loop 1 is isothermal with Log T $=$ 5.85, both along the line of 
sight as well as along the length of the loop leg that is visible in the CDS field of view. Loop 2 is 
hotter than Loop 1. It is multithermal along the line of sight, with significant emission between 
6.2 $<$ Log T $<$ 6.4, but the loop apex region is out of the CDS field of view so it is not possible 
to determine the temperature distribution as a function of loop height. Loop 3 also appears to be 
multithermal, but a blended loop that is just barely resolved with CDS may be adding cool emission 
to the Loop 3 intensities and complicating our results. So, are coronal loops isothermal or multithermal? 
The answer appears to be yes!
\end{abstract}


\section{Introduction}

Standard models for quiescent coronal loops (e.g., Rosner, Tucker \& Vaiana 1978) assumed a 
single magnetic flux tube where the plasma was in hydrostatic equilibrium and had a uniform 
cross section. These models predicted a set of well-known scaling laws and a significant loop 
temperature gradient. Actual measurements of temperature and density are vital for understanding 
the physics of loops, and observations by the Coronal Diagnostics Spectrometer (CDS) on SOHO 
have produced a variety of results. For example, the loops analyzed by Del Zanna (2003) and 
Del Zanna \& Mason (2003) were isothermal along the line of sight and had a temperature gradient 
along their length, but the one analyzed by Brkovic et al. (2002) was isothermal in both directions. 
The loop analyzed by Schmelz et al. (2001) and Schmelz \& Martens (2006) was multithermal 
along the line of sight, with a temperature distribution that increased as a function of loop height.

The Brkovic et al. result agrees with those obtained by Lenz et al. (1999) using TRACE data. This 
triggered a resurgence of modeling analysis as the observed loops had substantially smaller 
temperature gradients than expected. The Schmelz et al. results support nanoflare models and 
require us to refine our view of coronal loops. Throughout this paper, we refer to a {\it loop} as a 
coherent structure seen in an observation and a {\it strand} as a magnetic flux tube for which the 
heating and plasma properties have approximately uniform cross section. Nanoflare models 
(e.g., Klimchuk \& Cargill 2001) as well as these CDS results indicate that 
some or perhaps even many loops are composed of an unknown number of strands.

The data used in this analysis were acquired as part of SOHO JOP 146, designed to provide 
simultaneous TRACE and CDS observations of active region coronal loops and address the 
apparent controversy that arose from results like those discussed above. These results may, 
however, be related to the fact that different {\it types} of loops were observed, but only additional 
observations and analyses can confirm this. The observations are described in \S 2, data analysis 
in \S 3, results in \S 4, and conclusions in \S 5.

\section{Observations}

Cirtain et al. (2007) give details of the SOHO JOP 146 data set, including timelines, 
calibration, coalignment, as well as the initial stages of analysis. TRACE (Handy et al. 1999) 
is a Cassegrain 
telescope with a 30 cm aperture and a field of view of 8.5$'$ $\times$ 8.5$'$. It images the 
corona using three EUV passbands centered at 171, 195, and 284 \AA. There are also UV 
passbands that can image the photosphere and the transition region, but these data were not used 
in this analysis. TRACE has state-of-the-art pixels (0.5$''$) and a spatial resolution of about of 
1$''$. It is capable of producing one EUV image per minute.

CDS (Harrison et al. 1995) is a Wolter-Schwartschild II grazing incident telescope, with a full 
geometric area of 
289.28 cm$^2$ and an effective focal length of 257 cm.  A light beam is focused at an entrance slit 
and then split in two, with one component feeding into the Normal Incidence Spectrometer (NIS) 
while the other beam is directed to the Grazing Incidence Spectrometer (GIS). Since all data for 
this project were taken exclusively from the NIS component, the GIS will not be discussed 
further. NIS data is structured into cubes by scanning the slit across the Sun from West to East 
onto a 1024 $\times$ 1024 pixel detector. The instrument has 1.68$''$ pixels and a spatial 
resolution of about 5-6$''$. These observations used the 4$''$ slit. For each slit position, two 
spectra are 
generated at each pixel, the first covering the wavelength range from 310 to 380 \AA, and the 
second covering 517 to 633 \AA. Thus a unique spectrum exists for each scan position and each 
pixel along the slit. NIS rarely transmits data from the entire wavelength range back to Earth due 
to telemetry limitations. Instead, small wavelength bands, usually centered on strong emission 
lines, are transmitted. 

Two types of CDS rasters were taken as part of SOHO JOP 146: context rasters and sparse 
rasters. The 4$''$ $\times$ 240$''$ slit was used in both cases. The context raster field of view 
was 240$''$ $\times$ 240$''$. It was composed of 60 exposures, each with duration of 60 s for a 
total accumulation time of approximately 85 minutes. For the sparse rasters, exposures were not 
taken at every possible slit position, but instead with 12$''$ or 16$''$ steps in between each 
exposure position. This significantly reduced the time required to complete a raster. The total 
field of view was 102$''$ $\times$ 240$''$, and data were collected at fixed positions on the solar 
disk every 7 minutes. 

The data analyzed in this paper were taken on 2003 January 17 of AR 10250. The start time for 
the CDS observations was 06:51:27 UT. The observing sequences ran for 7 hours and the context 
rasters were interspersed with the sparse rasters. TRACE images were collected in all three EUV 
filters. The 171 and 195 \AA\ images were taken at full resolution. The 284 \AA\ images were 
binned to decrease the exposure time, but unfortunately, were not useful for the present analysis.

\section{Analysis}

Some of the JOP 146 data have been analyzed by Cirtain et al. (2007) and Schmelz et al. 
(2007). Cirtain et al. (2007) used an emission measure (EM) loci technique (Jordan at al. 1987) 
to determine where and when the background-subtracted loop plasma was isothermal. Their 
analysis used many CDS lines observed in 26 sparse rasters taken from 10:43 to 16:45 UT on 
2003 January 17. They found that about half of the pixels investigated were consistent with 
isothermal plasma, but that these results change with both space and time. Some (but not all) 
pixels along the loop were isothermal some of the time.

Schmelz et al. (2007) used this data set to investigate various concerns related to the temperature 
analysis of CDS data as well as the inherent assumptions made in differential emission measure 
(DEM) techniques. They used the CDS context raster from 14:24 UT to address the relatively 
slow cadence and large pixel size of the CDS instrument, the inherent smoothing required in all 
automatic inversion DEM techniques, as well as limb vs. disk solar loop observations. They 
found that CDS DEM analysis was reliable and provided meaningful physical results for coronal 
loops. This paper is an extension of these two previous analyses. 

In coronal equilibrium, where upward transitions result from collisions between electrons and ions and downward transitions are
spontaneous, the intensity $I$ of an optically thin spectral line of wavelength $\lambda_{ij}$  can be written as

\begin{equation}
I(\lambda_{ij}) = {1 \over 4 \pi}\ {hc \over \lambda_{ij}}\ A\
\int_0^L G(T, n_e)\ n_e^2\ dl,
\end{equation}

\noindent
where $h$ is PlanckÕs constant, $c$ is the speed of light, $A$ is the elemental abundance, $T$ is 
the electron temperature, $G(T, n_e)$ is the contribution function, $n_e$ is the electron 
density, $l$ is the line-of-sight element, and $L$ is the depth of the observing region. If the 
emitting plasma can be characterized by a single temperature and density, the contribution function 
is taken out of the integral, which is then referred to as the emission measure (EM), the amount of 
emitting material along the line of sight:

\begin{equation}
I(\lambda_{ij}) = {1 \over 4 \pi}\ {hc \over \lambda_{ij}}\ A\ G(T, n_e)\ EM
\end{equation}

The EM loci method uses the observed intensity of each spectral line and plots the EM as a 
function of temperature (for different values of the electron density). If all the resulting curves intersect at a single point (within the 
uncertainties), then the plasma may be isothermal and the approximation is justified. If, 
however, the curves do not intersect, then the plasma is multithermal and the analysis becomes 
more complex. In this case we need to define the DEM, $\Phi(T)$, the amount of emitting 
material as a function of temperature along the line of sight:

\begin{equation}
\int_0^L G(T, n_e)\ n_e^2\ dl = \int_T G(T, n_e)\ \Phi(T)\ dT
\end{equation}

and 

\begin{equation}
I(\lambda_{ij}) = {1 \over 4 \pi}\ {hc \over \lambda_{ij}}\ A \int_T \
G(T, n_e)\ \Phi (T)\ dT
\end{equation}

The DEM problem is mathematically ill-posed so small changes in the observed intensities can 
produce large variations in the DEM curves, and a vast range of solutions can reproduce the 
spectra within arbitrarily small observational errors. As a result, DEM solutions have been 
criticized severely in the literature and rightly so (see, e.g., Judge \& McIntosh 1999). 
Enforcing a positive 
solution helps a bit, and no one disputes the fact that although a purely mathematical solution can 
have negative values, a DEM curve representing astrophysical plasma must be strictly positive. 
Many DEM methods also rely on smoothing. Again, in a purely mathematical solution, adjacent 
temperature bins can be independent of each other, but these solutions might not be physically 
meaningful. Different smoothing methods, however, often result in different DEM results. For an 
detailed discussion of DEM uncertainties and smoothing, please see Kashyap \& Drake (1998).

To get around these problems, authors sometimes impose ad hoc assumptions like the overly 
simplistic two-Gaussian temperature distribution for the loop strands modeled by Aschwanden et 
al. (2007). Artificial limits on the DEM temperature range are also employed when the resulting 
DEM curves are not constrained by the data itself. Excluding temperature bins, however, can 
artificially limit the range of possible solutions. The complicating factor is the intrinsically high 
dynamic range of the DEM curves. A suitably large combination of emissivity and DEM can 
easily overcome the small instrument response. Del Zanna, Bromage \& Mason (2003) describe 
a good example of this analysis for polar plumes. Schmelz, Kashyap \& Weber (2007) show just 
how large the DEM uncertainties can be for TRACE loop data when no ad hoc assumptions are 
imposed. These results caution against such analyses as they tend to over interpret the data.

This paper uses two methods to evaluate the DEM based on the CDS
data. The first uses the forward folding technique with a manual
manipulation of the curve as described by Schmelz et al. (1999). Although this method is time-consuming, it 
forces the user to understand both the limitations on the data and the assumptions going into the 
analysis. The user has control of the final DEM shape and no smoothing is required beyond that 
imposed by the resolution of the contribution functions (0.1 dex). The second method uses the 
automatic inversion technique 
described by Warren (2005) and Brooks \& Warren (2006). The DEM curve is represented with 
a series of spline knots that are repositioned interactively for more control over the smoothness 
of the DEM curve. This method represents the best of both worlds: the quickness of automatic 
inversion and the control of manual manipulation. In both cases, the best fit is determined from a 
$\chi^2$ minimization of the differences between the intensities observed by CDS and those predicted by Eq. 4. 

\section{Results}

Our results used the context rasters from 14:24 UT mentioned in the previous section to analyze 
three loops which were part of AR 10250. The emission lines used in this study are listed in Table 1. 
These range from Mg VI at 349.2 \AA\ with a peak formation temperature of Log T $=$ 5.6 to 
Fe XIX at 592.2 \AA\ with a peak formation temperature of Log T $=$ 6.80, although no photons were
detected at these high temperatures. Figure 1 shows the CDS field of view in a series of 
spectral lines with different peak formation temperatures. Figure 2 shows the pixels along each 
loop as well as the background pixels chosen for detailed analysis. We used version 5.1 of the 
CHIANTI atomic physics database (Dere et al. 1997; Landi et al. 2006) and the hybrid coronal 
abundances (Fludra \& Schmelz 1999). 

\subsection{An Isothermal Loop}

We begin our analysis with the loop seen in Figure 2a. The footpoint area of this loop is 
visible in Mg VI at 349.2 \AA\ with a peak formation temperature of Log T $=$ 5.6, but the 
main leg of the loop is best seen in Mg IX at 368.1 \AA\  with a peak formation temperature of 
Log T $=$ 6.0, and fades in lines with higher temperatures. We chose six pixels along the loop 
for detailed analysis. The CDS temporal resolution is relatively slow and there is often 
concern that the target loop could be evolving while the CDS slit steps across the region. This is 
not a problem for Loop 1, however, because it is orientated vertically, that is, along the CDS slit. 
In this configuration, all the pixels analyzed for Loop 1 were observed at about the same time 
and can therefore be compared directly with each other. 

Proper background subtraction is an essential step in the analysis, and a lot of work was done 
investigating potential background pixels or collection of pixels. The background structure varies 
from position to position and from spectral line to spectral line. A potential background pixel 
might appear dark at one wavelength and not at another. There are also loops of varying 
temperatures close to each other so cuts across the loops do not reveal the true background. 
These cuts intersect other loops which are visible in maps made in different spectral lines with 
different peak formation temperatures. That is why CDS is such a valuable tool for this type of 
analysis.

The best loop-background contrast for Loop 1 (in Mg IX) was about 2.7:1, so we knew background 
subtraction was going to be challenging. The crowded field added to the difficulties.
In our original analysis, we selected an individual background pixel for each loop pixel. Each 
pixel pair was at approximately the same y-coordinate, but the background pixel was about 10 
pixels to the right of the loop pixel. This method gave us results that seemed inconsistent with 
the loop appearance: we got significant values for background subtracted intensities for lines 
where the loop appeared invisible. We then modified our original approach and followed the 
method of Cirtain et al. (2007). We used a single background pixel for loop pixels A-D, which 
appear to have a background with minimum structure, and a second background pixel for loop 
pixels E-F, which appear to be superimposed on a background of moss. This approach subtracted 
too much flux from several of the pixels, and we finally settled on a single background pixel (see 
Figure 2). This method gave us the results that were most consistent with the loop appearance in 
the different spectral lines.

Our results for Loop 1 are shown in Figure 3. The left column shows the EM 
loci plots for each pixel, starting at the footpoint and moving up the loop leg (which corresponds 
to pixels with lower y-values in Figures 1 and 2). These plots can be used to determine if the 
background-subtracted loop plasma is isothermal. The plots in the middle column show the DEM 
analysis done with forward folding and a manual manipulation of the curve while those in the 
right column show the DEM analysis done with automatic inversion.

The EM loci plots and the forward-folding DEM curves indicate that the loop plasma at these 
positions is isothermal with Log T $=$ 5.85. The automatic-inversion DEM curves are a bit 
broader, but have the same peak temperature. The discrepancies are due in part to the inherent 
differences in the methods, but also to the smoothing incorporated into the automatic inversion. 
Our results confirm those of Schmelz et al (2007) who showed that the forward-folding method 
adds no artificial width to the DEM curve beyond the temperature resolution element imposed by 
the steps taken in the spectral line response functions (Log T $=$ 0.1). However the automatic 
inversion DEM method produces a natural temperature width related to the inherent smoothing 
associated with the inversion, so it is not possible to obtain a result that represents a truly 
isothermal plasma.

With these high quality CDS data, the main sources of uncertainty on the DEMs are the background 
subtraction, which we address in detail above, and the atomic data. In most cases, these uncertainties 
will dominate the line intensity error. A detailed discussion of atomic data is beyond the scope of this 
paper, and although great progress has been made in recent years, there are still challenges. We have
tried to focus on reasonably strong lines which are well-isolated spectrally, but unidentified blends
may still be present. In addition, atomic physicists generally quote uncertainties of $\pm$20\%, but
atomic data for some lines are likely to be worse. These potential errors are almost never accounted 
for in DEM analysis and in certain cases, may dominate the DEM uncertainties.

It is unfortunate that there are no density-sensitive ion pairs remaining after background 
subtraction. Figure 4 plots what information we have about density. It shows the predicted (from 
forward folding) to observed intensity ratios for densities of (a) 10$^8$, (b) 10$^9$, and 
(c) 10$^{10}$ cm$^{-3}$. We did this analysis for all the pixels and show a typical example. 
We chose to display the results for pixel C because it was far enough away from the footpoint 
and in the middle of the distribution of loop pixels. Unfortunately, most of the lines are not 
density sensitive in this range, but the Si X line indicates that the plasma density is unlikely to 
be as low as 10$^8$ cm$^{-3}$, but 10$^9$ and 10$^{10}$ look equally good. Since the loop 
apex is well out of the CDS field of view, it is not possible to say with certainty that this loop is 
over-dense, but assuming that the apex temperature is the same as that for the positions
analyzed here, then the loop is indeed over-dense with respect to results expected from scaling 
laws (see, e.g., Warren \& Winebarger 2003).

The isothermal results for Loop 1 are consistent with the TRACE data (Figure 5). The flux from the 
ensemble of TRACE pixels corresponding to the CDS background pixel is subtracted from the 
flux of the ensembles for the various loop pixels in both the 195 and 171 TRACE images. 
We propagated the TRACE uncertainties through the pixel averaging process, background subtraction, 
and 195/171 division. We found that the TRACE ratios for the various Loop 1 pixels were not statistically different. 
The results are shown in Figure 6, which plots the 195/171 response as a function of Log T. 
The solid flat lines represent the flux ratios for the background subtracted pixels, 
where the spread is a measure of the uncertainty. The 
intersections are very close to a local minimum in the response ratio, and all the results are 
consistent with the CDS temperatures of Log T $=$ 5.85. This shows that the TRACE image ratio 
technique works well if the plasma is isothermal, but please see the next two subsections for 
different results (the dashed lines are for Loop 3).

\subsection{A Multithermal Loop}

We continue our analysis with the loop in Figure 2b (right). The footpoint area of this loop is 
visible in almost every spectral line in the data set. This indicates to us that perhaps many 
individual loops with different temperatures or temperature distributions are anchored at this 
position. As we move up the leg of the loop, we see that it remains visible in all the hotter lines 
(Log T $>$ 6.1), but is not visible in the cooler lines. It is most distinct in Si XII at 520.7 \AA\ 
with a peak formation temperature of Log T $=$ 6.3, and is still visible in the two Fe XVI lines, 
but not in Fe XIX where a significant intensity would most likely indicate flaring plasma.
As with Loop 1, we chose several pixels along the loop for detailed analysis. Note that Loop 2 is 
also orientated along the CDS slit, so all the pixels were observed at about the same time 
and can be compared directly with each other. 

The best loop-background contrast for Loop 2 (in Si XII) was about 2:1, so background subtraction 
was even more challenging than it was for Loop 1.  We first 
chose the same background pixel, but all the spectral lines, including Si XII where the loop was 
clearly visible, had insignificant fluxes after background subtraction. We had subtracted 
too much flux. We experimented with different individual pixels and with the pixel-pair method 
that had not worked with Loop 1. We settled on a single pixel (see Figure 2b - right); as with Loop 1, 
this approach seemed the most consistent with the loop appearance in the different spectral lines.

Our results for Loop 2 are shown in Figure 7. The format is the same as that of Figure 3; from 
left to right, we have the EM loci plots for each pixel, the forward-folding DEM results, and the 
automatic-inversion DEM results. These results are different than those for Loop 1: the curves 
for the different spectral lines do not intersect at a single point in the EM loci plots, indicating 
that the background-subtracted loop plasma is not isothermal. The two DEM methods agree 
(with this result and with each other). A multi-thermal distribution is required to reproduce the 
observed spectral line intensities. As mentioned above, the footpoint pixel seen in (A) may be
an area where many individual loops with different temperatures or temperature distributions 
are anchored. This could account for the very broad distribution. For the loop-leg pixels (B-F), 
the cool temperature end of the DEM distributions are strongly 
constrained since the loop is not seen in any lines with peak 
formation temperature log T $<$ 6.15. The high-temperature constraint is somewhat weaker. 
The loop is visible in Fe XVI (Log T $=$ 6.40) but not in Fe XIX (Log T $=$ 6.80), so the DEM 
curves could be a bit wider than those shown in Figure 7, but not narrower. 

This broad DEM result is reminiscent of the analysis of a loop on the limb observed by CDS on 
1998 April 20 and analyzed by Schmelz et al. (2001) and Schmelz \& Martens (2006). That loop 
was also most distinct in the Si XII line at 520.7 \AA, but there was concern that the broad DEM 
might be the result of the long path length for a loop on the limb. Loop 2 appears to have similar 
properties, but as is obvious from the images in Figure 1, it is situated on the solar disk. Of 
course it is still possible that the broad DEM distributions seen in Figure 7 are the result of loops 
of different temperatures along the line of sight and unresolved by the moderate pixel size of 
CDS, but these distributions could also result from a single loop composed of isothermal strands 
of different temperatures.

Our density analysis for Loop 2 is shown in Figure 8. The format is similar to that of Figure 4 for 
Loop 1, showing the predicted (from forward folding) to observed intensity ratios for the spectral 
lines detected in pixel ÔCÕ for densities of (a) 10$^8$, (b) 10$^9$, and (c) 10$^{10}$ cm$^{-
3}$. The two coolest lines (Al XI and Fe XII) and the three hottest lines (Si XII and the two Fe XVI 
lines) are not sensitive to density in this range, but the intermediate lines (Fe XIII and the two 
Fe XIV lines) are. These lines indicate that the plasma density is unlikely to be as high as 
10$^{10}$ cm$^{-3}$ or as low as 10$^8$ cm$^{-3}$. The best-fit density was 
4 $\times$ 10$^8$ cm$^{-3}$.

Unfortunately, Loop 2 does not show up in the TRACE images. This is simply a matter of bad timing. 
During the lifetime of Loop 2, TRACE was imaging with the 171 and 284 \AA\ channels. The response 
of the former is too cool and the confusion in the latter for this particular active region was too high.

\subsection{A Curved Loop}

The last loop selected for analysis is shown in Figure 2b (left). This loop is seen in many spectral lines 
with a wide range of temperatures and it is the most distinct loop structure in the TRACE image. 
The best loop-background contrast for Loop 3 (in Si XII) was about 2.3:1, and background subtraction 
was just as challenging as it was for Loops 1 and 2. As before, we eventually 
settled on a single pixel because this approach seemed the most consistent with 
the loop appearance.

Our results for Loop 3 are shown in Figures 9 and 10. The format is the same as that of Figures 3 
and 7. The EM loci plots show that the curves for the different spectral lines do not intersect at a 
single point, indicating that the background-subtracted loop plasma is not isothermal and that a 
multi-thermal distribution is required. However, we found two different DEM models that could 
reproduce the observed spectral line intensities. One is very broad (Figure 9) and the other has 
two more narrow components (Figure 10). This result shows that the DEM loop analysis is not 
always clean and that in this particular case, the CDS data are not good enough to distinguish 
between these models. 

Our density analysis for Loop 3 is shown in Figure 11. The format is similar to that of Figures 4 
and 8, showing the predicted (from forward folding) to observed intensity ratios for the spectral 
lines observed in pixel ÔCÕ for densities of (a) 10$^8$, (b) 10$^9$, and (c) 10$^{10}$ cm$^{-
3}$. As with Loop 2, these lines indicate that the plasma density is unlikely to be as high as 
10$^{10}$ cm$^{-3}$ or as low as 10$^8$ cm$^{-3}$, but 10$^9$ cm$^{-3}$ looks quite good. 
The best-fit density was 3 $\times$ 10$^9$ cm$^{-3}$.

The TRACE analysis for Loop 3 is shown in Figure 6. As with Loop 1, the flux from the 
ensemble of TRACE pixels corresponding to the CDS background pixel is subtracted from the 
flux of the ensembles for the various loop pixels in both the 195 and 171 TRACE images.
The dashed lines represent the flux ratios for the background subtracted pixels, and the 
intersections with the response ratio range from T $=$ 1.1$-$1.2 MK. This result may appear 
inconsistent with that of CDS, but a possible explanation was provided by Weber et al.\ (2005). 
If an isothermal approximation is used when the 
plasma is actually multithermal, then the resulting 171-to-195-\AA\ ratio will be 0.81. This value 
corresponds to the ratio of the areas under the 171- and 195-\AA\ response curves and implies a 
plasma temperature of 1.2 MK in the isothermal approximation. In other words, a TRACE 
temperature of 1.2 MK is consistent with multithermal plasma, and therefore, with the results 
obtained with CDS.

\section{Conclusions}

We have analyzed three different loops that were observed with CDS and TRACE and obtained 
three different results. EM loci plots and DEM with forward folding indicate that the loop plasma 
at each chosen pixel of Loop 1 appears to be isothermal with Log T $=$ 5.85. DEM curves made 
with automatic inversion (Figure 3) and TRACE 195-to-171 image ratios (Figure 6) are 
consistent with this result. We also find that all the pixels along the leg of Loop 1 appear to have 
the same temperature, a result consistent with image ratio analysis (e.g., Neupert et al. 1998; 
Lenz et al. 1999) and inconsistent with results predicted by standard models for loops in 
hydrodynamic equilibrium (e.g., Rosner, Tucker \& Vaiana 1978). Although CDS has 
moderate-size pixels, it has been used by other authors to find isothermal plasma in loops (e.g., 
Brkovic et al. 2002; Del Zanna 2003; Del Zanna \& Mason 2003), in polar plumes (Del 
Zanna, Bromage \& Mason 2003), and in an area of quiet Sun off the limb (Allen et al. 2000). 
The results for Loop 1 just add to this list.

Loop 2 gives a different result. DEM curves made with both forward folding and with automatic 
inversion indicate that the loop plasma at each pixel is multithermal. The EM loci plots are 
consistent with this result. This is similar to the conclusions of Schmelz et al.\ (2001) and 
Schmelz \& Martens (2006) for a loop observed on the solar limb with CDS. It may be a 
coincidence, but both of these loops were best seen in the Si XII line, one of hotter (but non-
flaring) lines available in the CDS wavelength range. This could be a filling-factor effect 
(Brickhouse \& Schmelz 2006), meaning that there is simply more material along the line of 
sight at Log T $=$ 6.3 than there is at Log T $=$ 5.85, or it could be a consequence of loop 
cooling (Winebarger et al. 2003). Patsourakos et al. (2002) showed that slightly different initial 
conditions for cooling loops in an arcade were sufficient to reproduce qualitatively the fuzzy 
appearance of EUV post-flare loops seen for example in the TRACE 284 \AA\ channel.

Results for Loop 3 are not as clear. The observed CDS intensities can be reproduced with a very 
broad DEM curve (Figure 9) or a two-component model (Figure 10). We favor the 
two-component model based on the CDS images. Figure 1c shows the active region structure in 
Ca X (Log T $=$ 5.85). There is a cool loop slightly offset from Loop 3, but not completely 
resolved with the CDS instrument. We suspect that cooler emission from this Ca X loop is 
blended with Loop 3 and contributing to the cooler DEM 
component seen in Figure 10. Subtracting the emission making up the cool component from the 
individual spectral line intensities (which we now assume come from the Ca X loop) leaves the 
Loop 3 intensities alone which can then be modeled with a single component (although a bit 
broader) at a temperature centered near Log T $=$ 6.3.

Although CDS has in many cases detected isothermal plasma, the moderate spatial resolution 
(5-6$''$) of the instrument has long been a concern for loop studies. The analysis of these three 
loops illustrates different strengths and weaknesses. The CDS observations clearly discerned the 
isothermal nature of Loop 1, even thought the field is crowded with other loops and plagued with 
a complicated background structure. The broader DEM components like the one required for 
Loop 2 need to be confirmed using data from spectrometers with better spatial resolution. 
Problems like the apparent blending of the Ca X and Loop 3 structures, which are not quite 
cospatial in the CDS images, will also be resolved with future instruments.

\acknowledgments
We would like to thank Harry Warren for the use of his DEM program, Amy Winebarger for 
installing this program on our computers, and UofM physics students James Andrews and 
Jennifer Garst for help with analysis. Solar physics research 
at UofM is supported by NSF ATM-0402729 and NASA NNG05GE68G. HEM and GDZ 
acknowledge financial support from PPARC. We are grateful to P.R. Young and the entire CDS 
team for their help in obtaining these observations. We also benefited greatly from discussions 
at the coronal loops workshops in Paris (November 2002), Palermo (September 2004), and 
Santorini (June 2007).

{}

\clearpage

\begin{deluxetable}{lllrcllll}
\tabletypesize{\scriptsize}
 \tablewidth{0pt}
\tablecaption{ CDS Spectral Lines}
\tablehead{
\colhead{$\lambda$ (\AA) } & \colhead{Ion } & \colhead{Log T } & \colhead{Atomic }&& \colhead{Transition } &\colhead{Locations}& \colhead{where line}& \colhead{ is present }
}
\startdata

349.164&  	Mg VI&  	5.60& 2s$^2$ 2p$^3$ $^2$D$_{5/2}$ &$-$&2s 2p$^4$ $^2$D$_{5/2}$ \\
558.592&	Ne VI& 	5.60& 2s$^2$	2p	$^2$P$_{1/2}$	&$-$&	2s	2p$^2$		$^2$D$_{3/2}$&	Loop 1		\\
562.798&	Ne VI&	5.60&	2s$^2$	2p	$^2$P$_{3/2}$	&$-$&2s	2p$^2$	 	$^2$D$_{5/2}$&	Loop 1		\\
557.766&	Ca X&		5.85&	3s		$^2$S$_{1/2}$	&$-$&3p	$^2$P$_{3/2}$&	Loop 1	&Ftpt 2	\\
368.070&	Mg IX&	6.00&	2s$^2$ 	$^1$S$_{0}$	&$-$&2s	2p	$^1$P$_{1}$&	Loop 1&	Ftpt 2	&Loop 3\\
349.860&	Si IX&		6.05&	2s$^2$	2p$^2$	$^3$P$_{2}$	&$-$&2s	2p$^3$	 $^3$D$_{1,2,3}$&	Loop 1&	Ftpt 2	&Loop 3\\
624.943&	Mg X&	6.05&	1s$^2$	2s	$^2$S$_{1/2}$	&$-$&1s$^2$	2p	$^2$P$_{1/2}$&	&	Ftpt 2	&Loop 3\\
347.409&	Si X&		6.15&	2s$^2$	2p	$^2$P$_{1/2}$	&$-$&2s	2p$^2$		$^2$D$_{3/2}$&	&	Ftpt 2	&Loop 3\\
356.030&	Si X&		6.15&	2s$^2$	2p	$^2$P$_{3/2}$	&$-$&2s	2p$^2$		$^2$D$_{3/2,5/2}$&	&	Ftpt 2	&Loop 3\\
346.853&	Fe XII&	6.15&	3s$^2$	3p$^3$	$^4$S$_{3/2}$	&$-$&3s	3p$^4$		$^4$P$_{1/2}$&	&	Ftpt 2	&Loop 3\\
352.107&	Fe XII&	6.15&	3s$^2$	3p$^3$	$^4$S$_{3/2}$	&$-$&3s	3p$^4$		$^4$P$_{3/2}$&	&	Ftpt 2	&Loop 3\\
364.468&	Fe XII&	6.15&	3s$^2$	3p$^3$	$^4$S$_{3/2}$	&$-$&3s	3p$^4$		$^4$P$_{5/2}$&	&	Loop 2	&Loop 3\\
348.184&	Fe XIII	&	6.20&	3s$^2$	3p$^2$	$^3$P$_{0}$	&$-$&3s	3p$^3$		$^3$D$_{1}$&	&	Loop 2	&Loop 3\\
550.032&	Al XI&	6.20&	1s$^2$	2s	$^2$S$_{1/2}$	&$-$&1s$^2$	2p	$^2$P$_{3/2}$&	&	Loop 2	&Loop 3\\
334.180&	Fe XIV&	6.30&	3s$^2$	 3p 	$^2$P$_{1/2}$	&$-$&3s	3p$^2$		$^2$D$_{3/2}$&	&	Loop 2	&Loop 3\\
353.837&	Fe XIV&	6.30&	3s$^2$	3p 	$^2$P$_{3/2}$	&$-$&3s	3p$^2$		$^2$D$_{5/2}$&	&	Loop 2	&Loop 3\\
520.666&	Si XII&	6.30&	1s$^2$	2s	$^2$S$_{1/2}$	&$-$&1s$^2$	2p	$^2$P$_{1/2}$&	&	Loop 2	&Loop 3\\
335.410&	Fe XVI&	6.40&	3s		$^2$S$_{1/2}$	&$-$&3p		$^2$P$_{3/2}$&	&	Loop 2	&Loop 3\\
360.759&	Fe XVI&	6.40&	3s		$^2$S$_{1/2}$	&$-$&3p		$^2$P$_{1/2}$&	&	Loop 2	&Loop 3\\
592.236 &Fe XIX&	6.80&	2s$^2$ 2p$^4$ $^3$P$_{2}$	&$-$& 2s$^2$ 2p$^4$ $^1$D$_2$
\enddata
\end{deluxetable}

\clearpage

\begin{figure}
\epsscale{.8}
\plotone{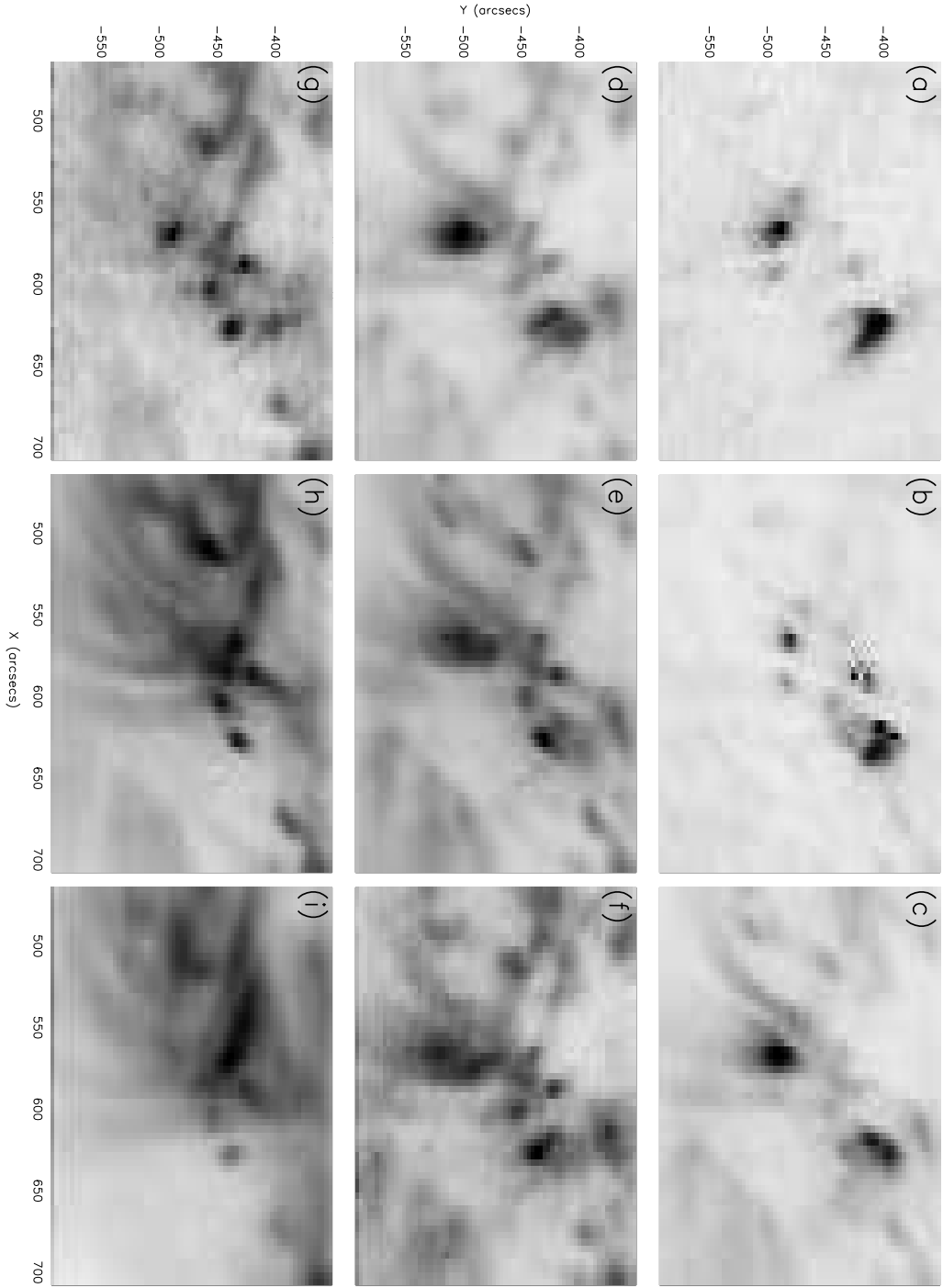}
%\plottwo{f1a.eps}{f1b.eps}
\caption{CDS images taken on 2003 January 17 at 06:51:27 UT of AR 10250 in order of increasing peak formation temperature:
(a) Mg VI (Log T $=$ 5.6); 
(b) Ne VI (Log T $=$ 5.6); 
(c) Ca X (Log T $=$ 5.85); 
(d) Mg IX (Log T $=$ 6.0); 
(e) Mg X (Log T $=$ 6.05); 
(f) Si X (Log T $=$ 6.15); 
(g) Fe XIV (Log T $=$ 6.3); 
(h) Si XII (Log T $=$ 6.3); and 
(i) Fe XVI (Log T $=$ 6.4). These are negative images.
}
\end{figure}

\clearpage

\begin{figure}
\epsscale{.5}
\plotone{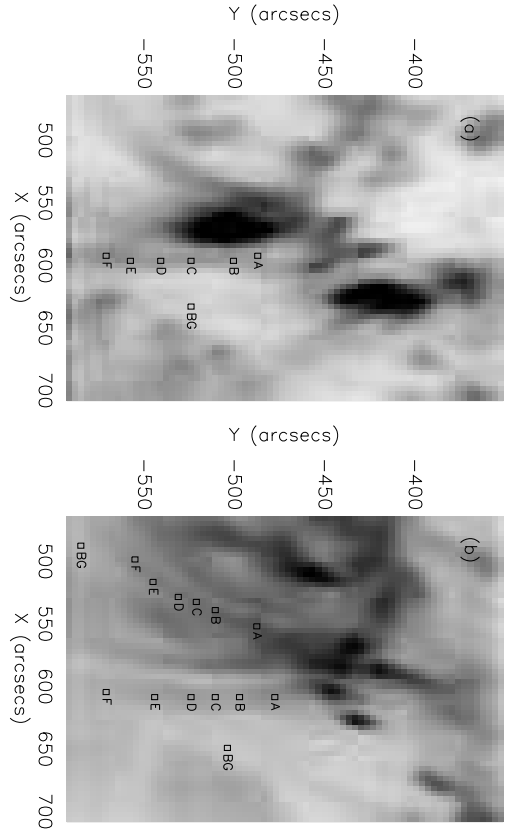}
%\plottwo{f1a.eps}{f1b.eps}
\caption{ A subset of the CDS images from Figure 1 showing the loop pixels and background pixels chosen for detailed analysis.
(a) CDS Mg IX at 368.1 \AA\ with Log T $=$ 6.0 showing the pixels for
Loop 1; 
(b) CDS Si XII at 520.7 \AA\ with Log T $=$ 6.3 showing the pixels for
Loop 2 (right) and Loop 3 (left). These are negative images.
}
\end{figure}

\clearpage

\begin{figure}
\epsscale{.9}
\plotone{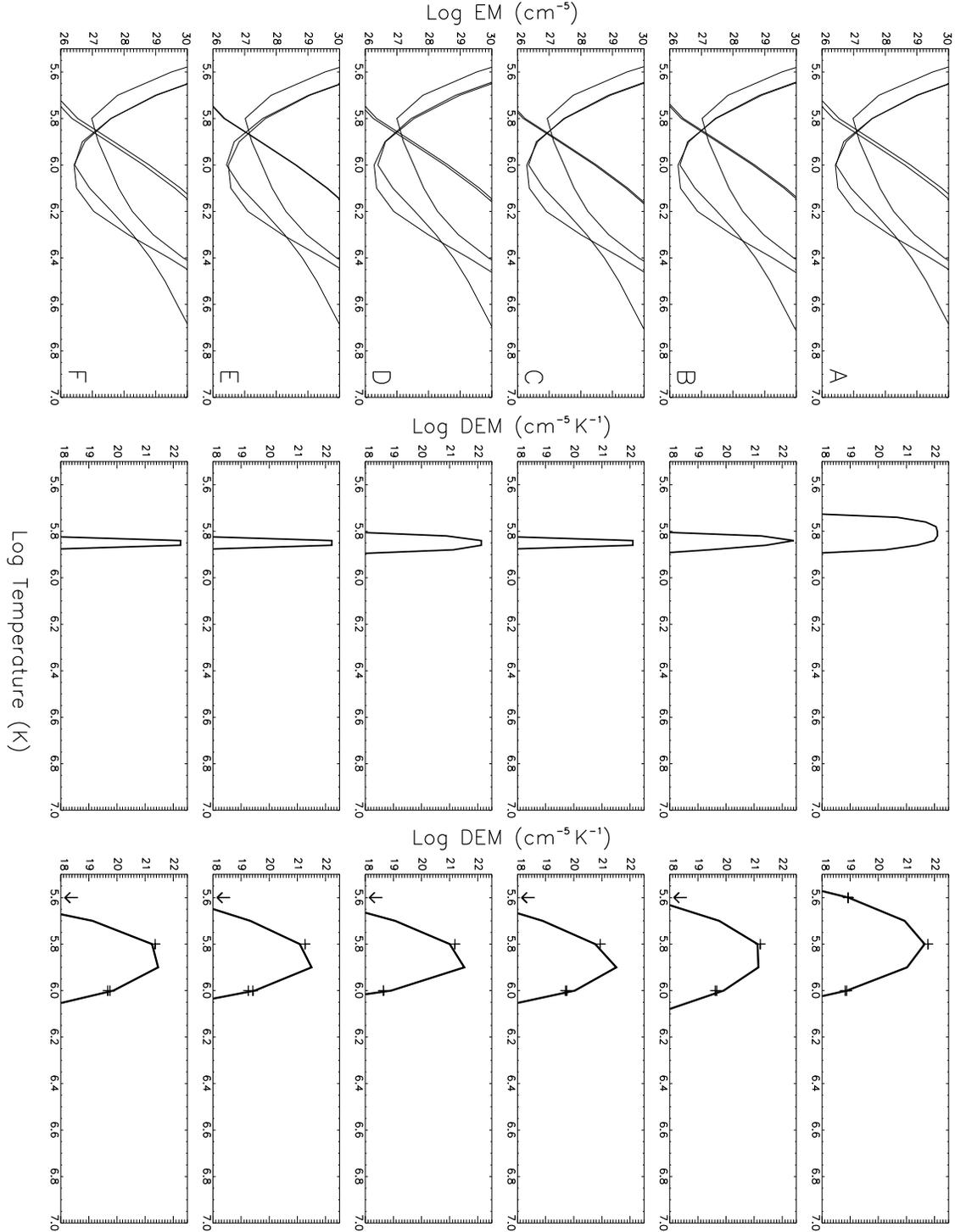}
\caption{The temperature results for the six pixels in Loop 1 after background subtraction, 
including EM loci plots (left), DEM curves with forward folding (middle), and DEM curves 
with automatic inversion (right). The plus signs represent the ratio of the observed to calculated 
radiance scaled by the emission measure and plotted at the peak formation temperature. 
Arrows indicate lines that fall below the y-axis minimum, but still give an excellent fit. 
(Note: these symbols are not included for the plots in the middle column because the spike DEM shape 
can fall between the data points, changing all the plus signs to arrows, and giving the 
incorrect impression of a bad fit.)
}
\end{figure}

\clearpage

\begin{figure}
\epsscale{.9}
\plotone{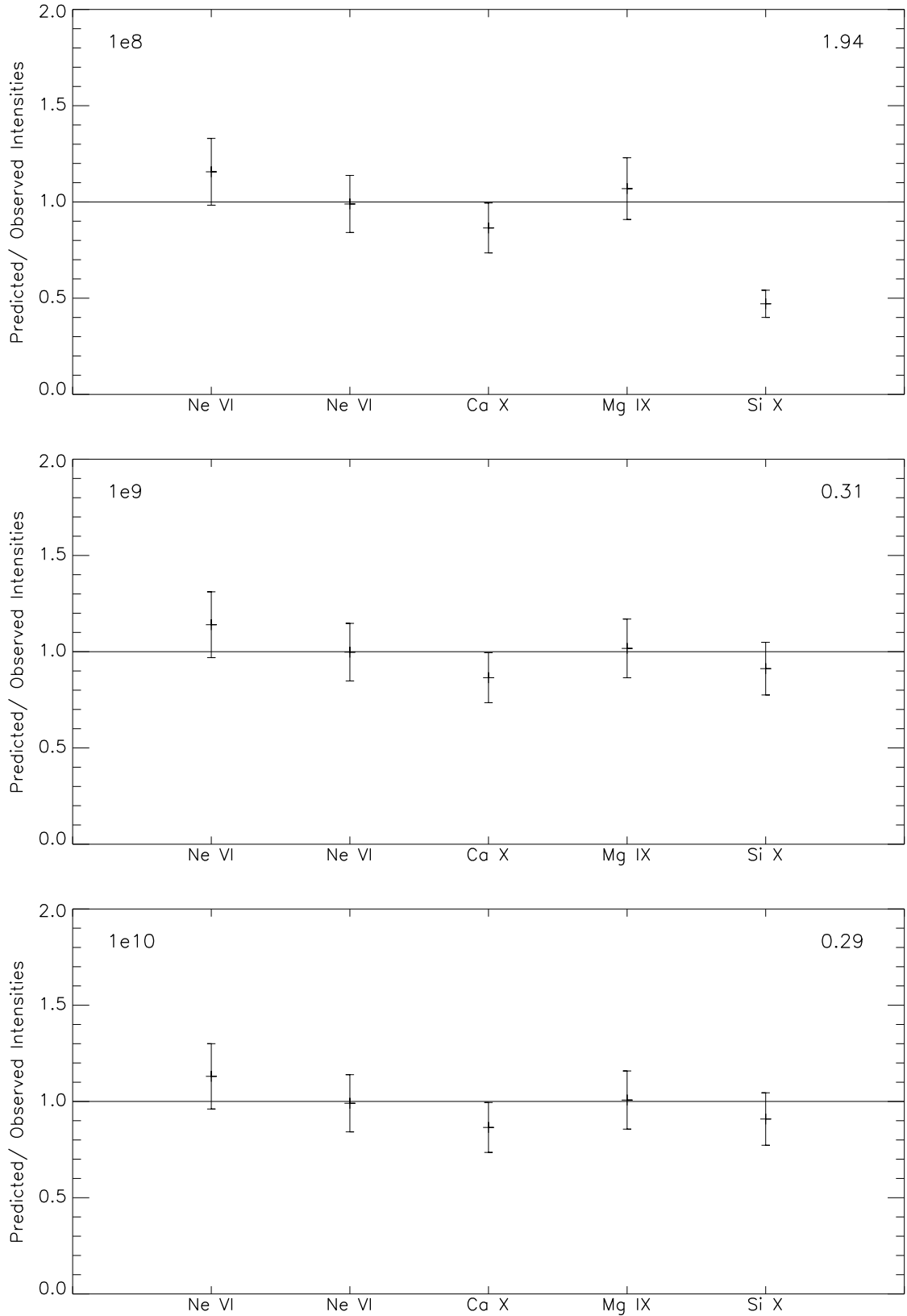}
\caption{Loop 1 density analysis. The predicted (from forward folding) to observed intensity ratios 
for the spectral lines observed in pixel ÔCÕ for (a) 10$^8$, (b) 10$^9$, and (c) 10$^{10}$ cm$^{-3}$. 
The number in the upper right corner of each panel is the reduced $\chi^2$.
}
\end{figure}

\clearpage

\begin{figure}
\epsscale{.85}
\plotone{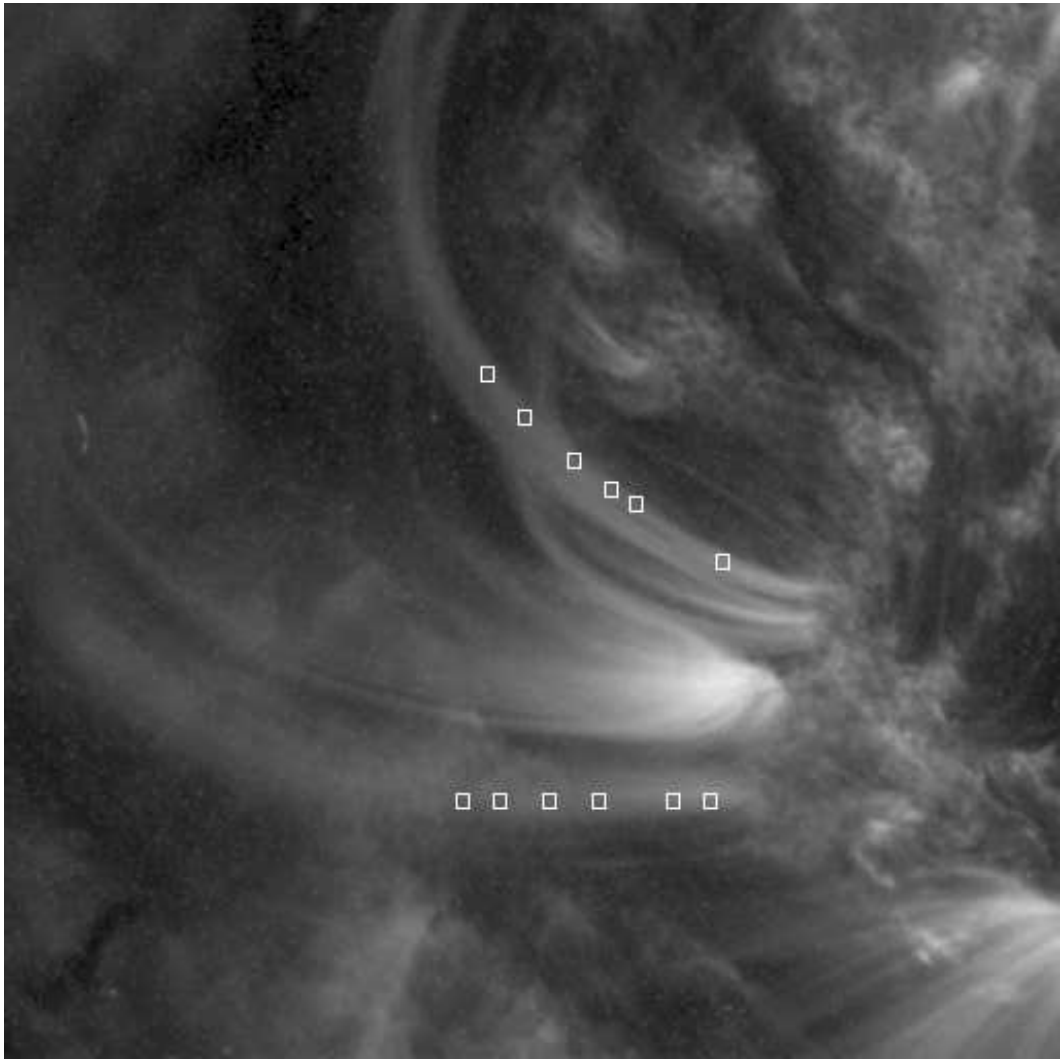}
\caption{TRACE 195-\AA\ image showing pixels chosen for detailed analysis for Loops 1 (left) and 3 (right). 
This is a negative image.
}
\end{figure}

\clearpage

\begin{figure}
\epsscale{.85}
\plotone{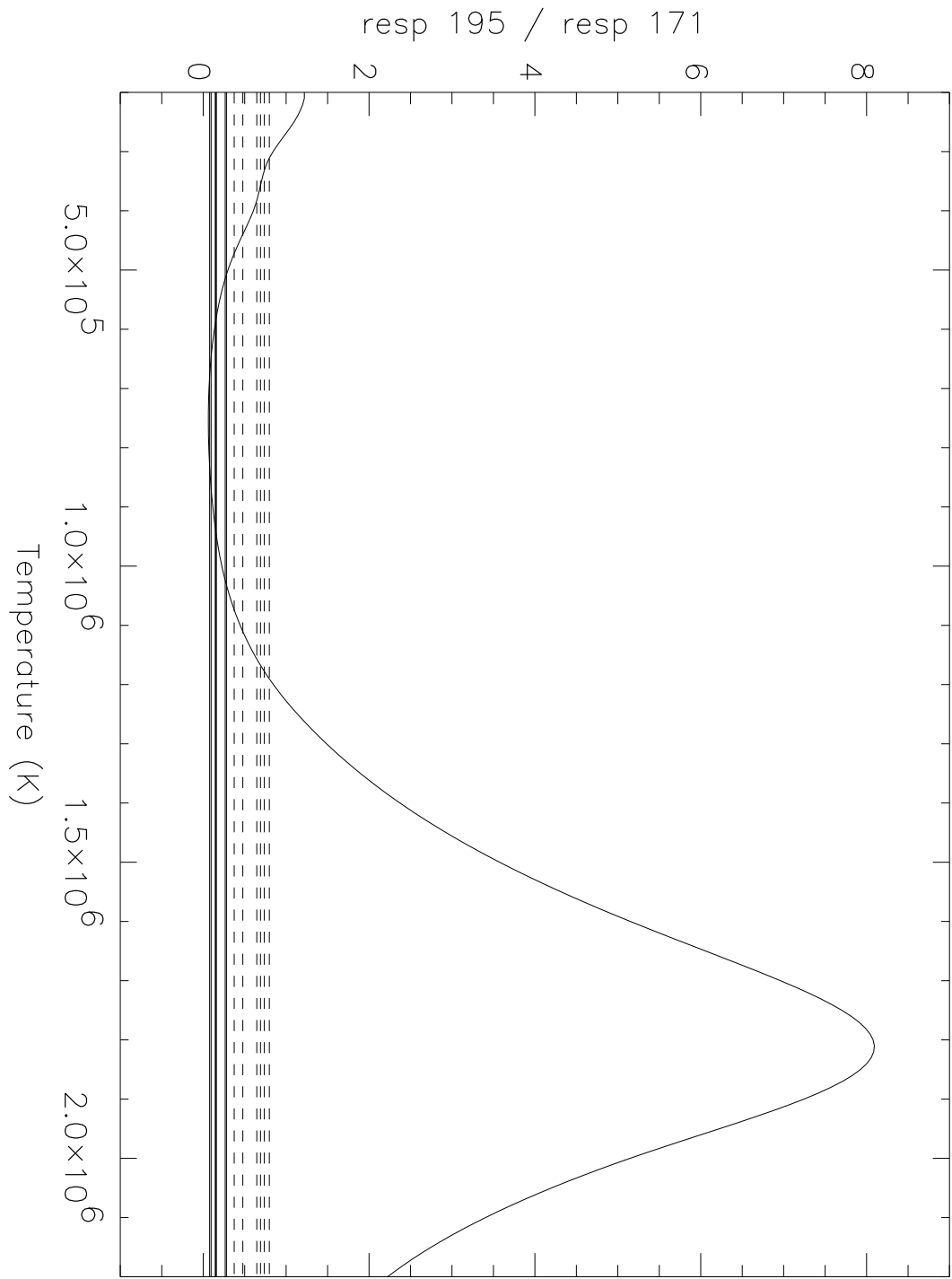}
\caption{TRACE 195/171 response function ratio vs. Log T. The flat lines represent the flux 
ratios for the background-subtracted pixels. The solid lines are for Loop 1 and the dashed lines 
are for Loop 3. The spread in each set is a measure of the uncertainty. The intersections of the  
solid lines are all close to a local minimum with T $=$ 0.7$\pm$0.2 MK (Log T $=$ 5.85), 
which is consistent with the isothermal CDS result for Loop 1. The intersections for the dashed 
lines are at T $=$ 1.1$-$1.2 MK, consistent with the multithermal CDS result for Loop 3 (see text).
}
\end{figure}

\clearpage

\begin{figure}
\epsscale{.9}
\plotone{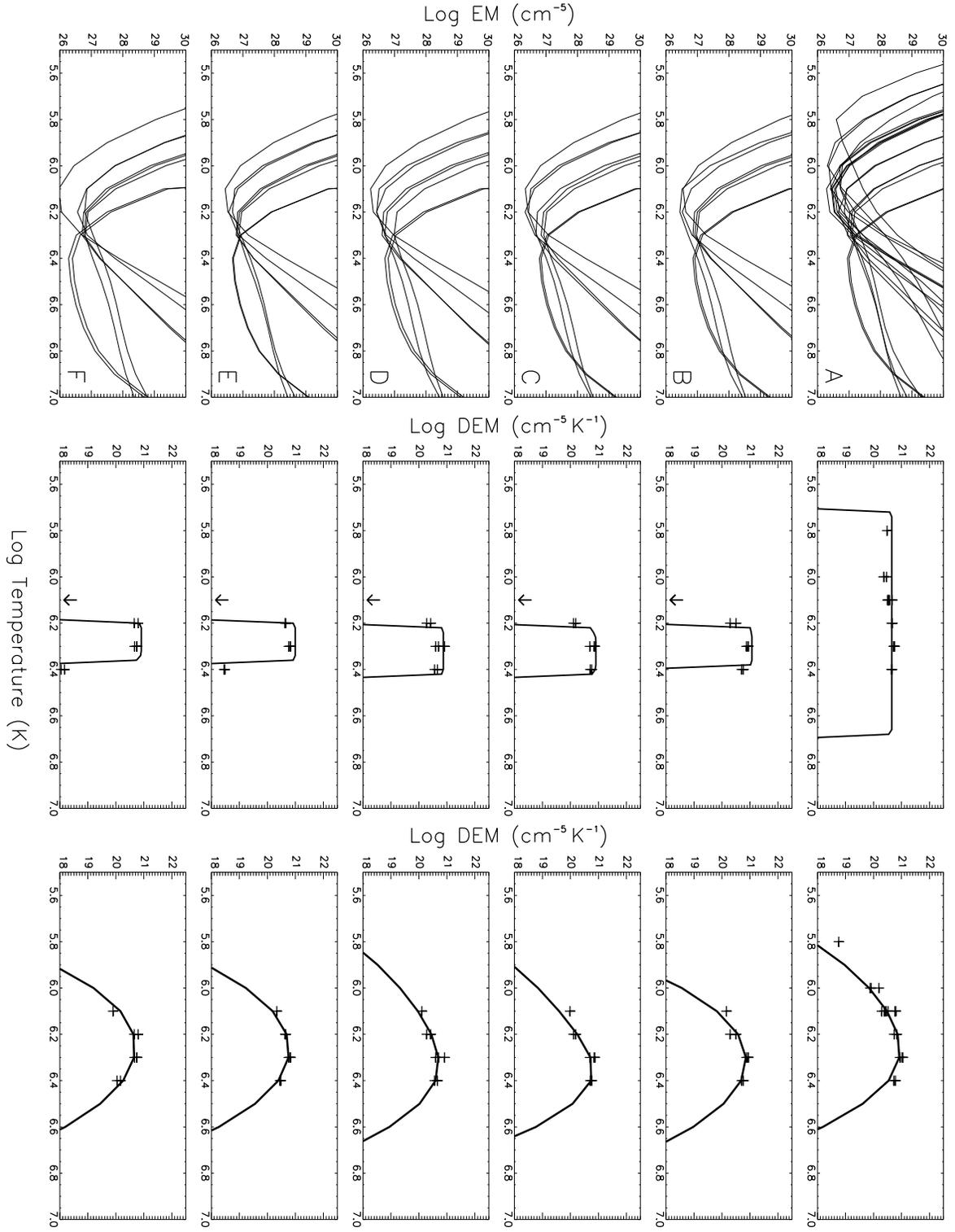}
\caption{Temperature analysis. Same as Figure 3, but for Loop 2.
}
\end{figure}

\clearpage

\begin{figure}
\epsscale{.9}
\plotone{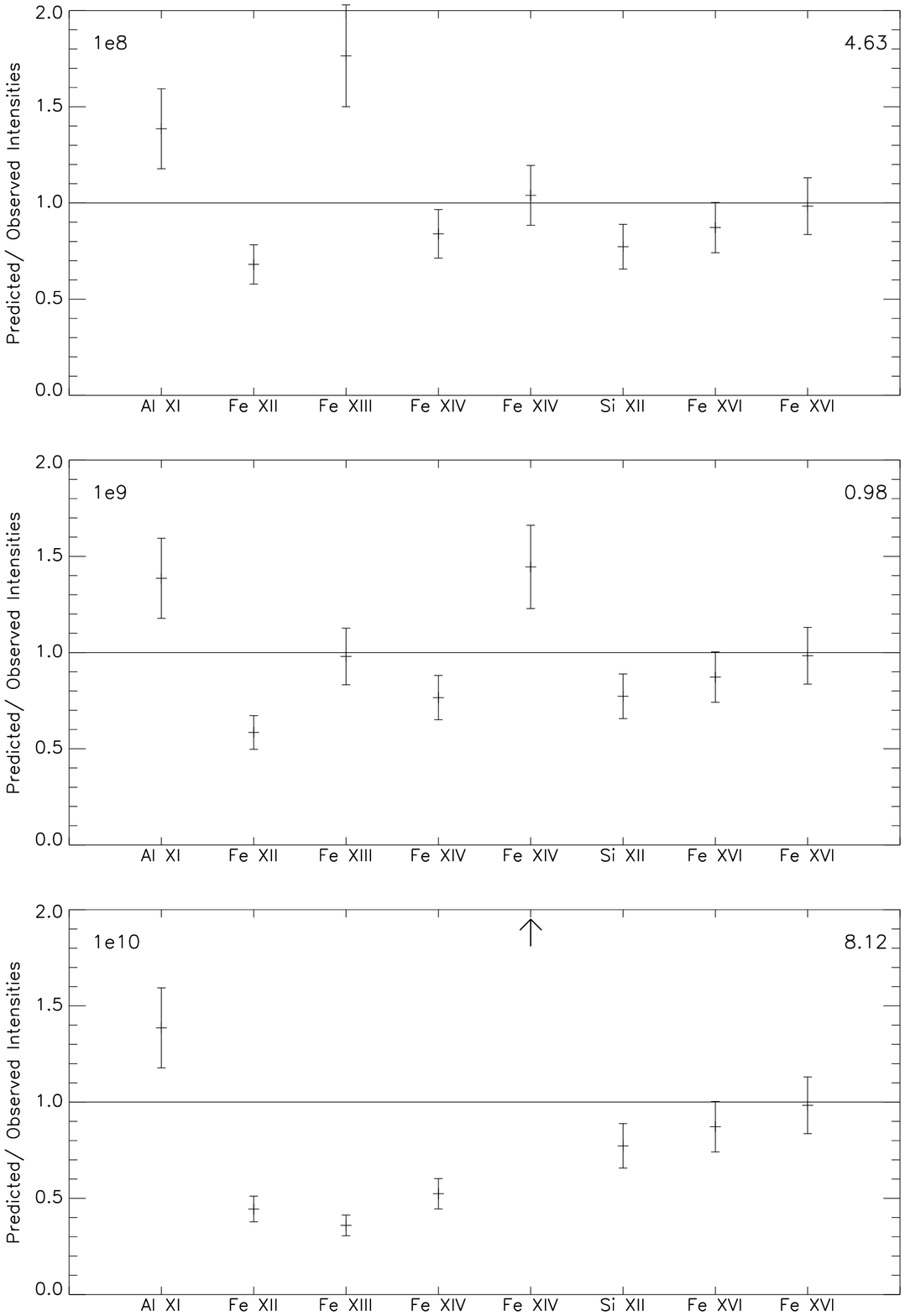}
\caption{Density analysis. Same as Figure 4, but for Loop 2.
}
\end{figure}

\clearpage

\begin{figure}
\epsscale{.9}
\plotone{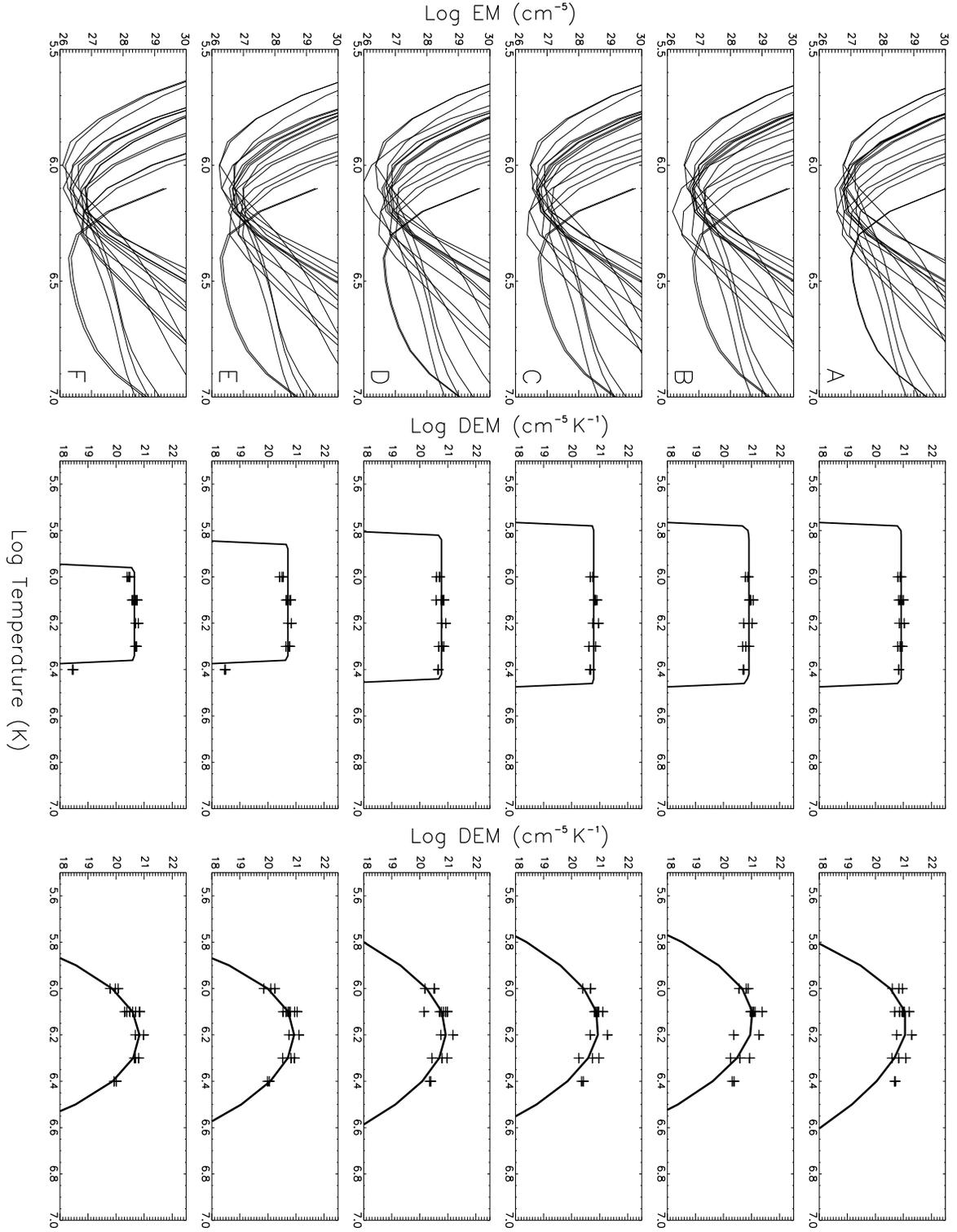}
\caption{Temperature analysis.  Same as Figure 3, but for Loop 3 (broad DEM model).
}
\end{figure}

\clearpage

\begin{figure}
\epsscale{.9}
\plotone{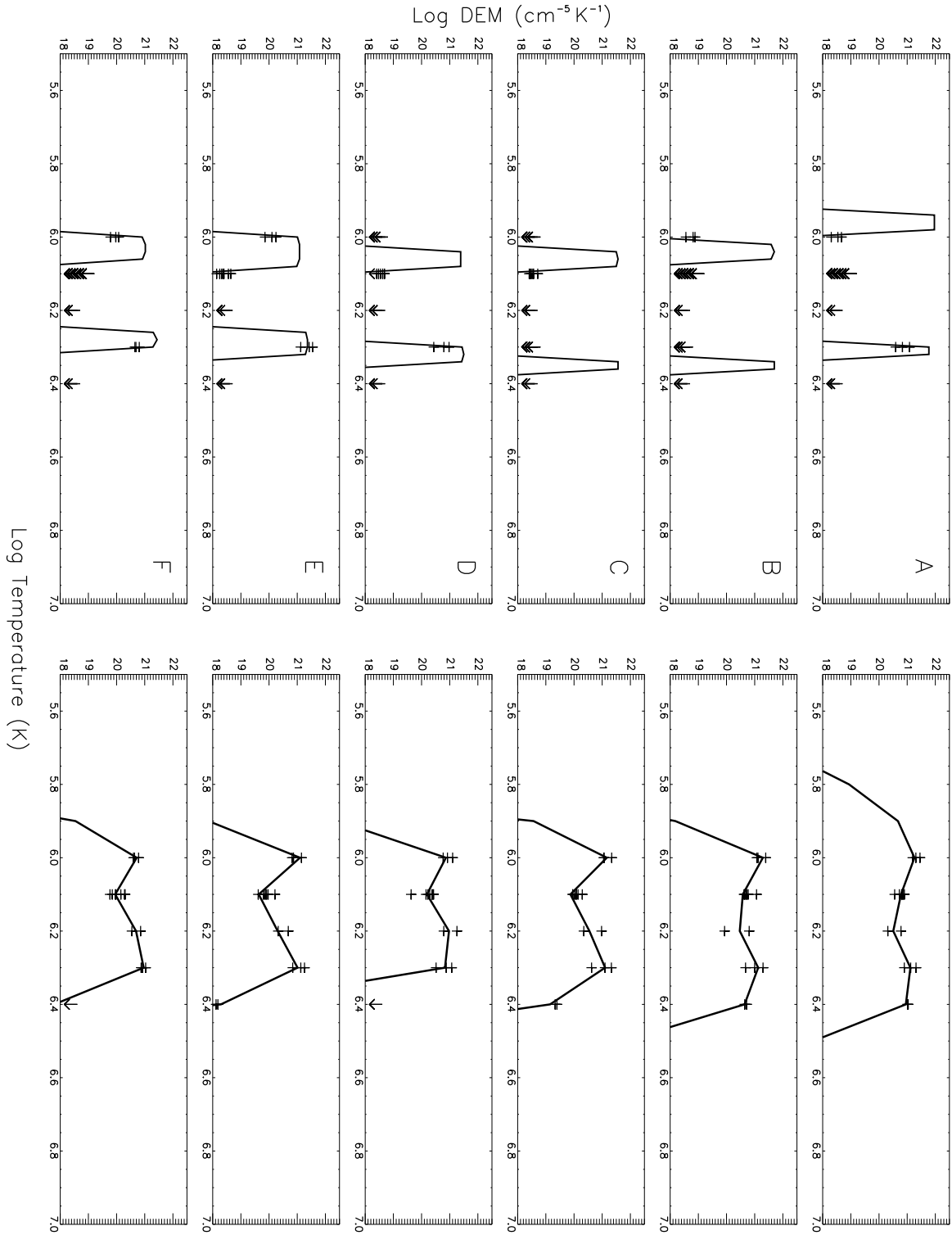}
\caption{Temperature analysis. Same as Figure 3, but for Loop 3 (two-component model).
}
\end{figure}

\clearpage

\begin{figure}
\epsscale{.9}
\plotone{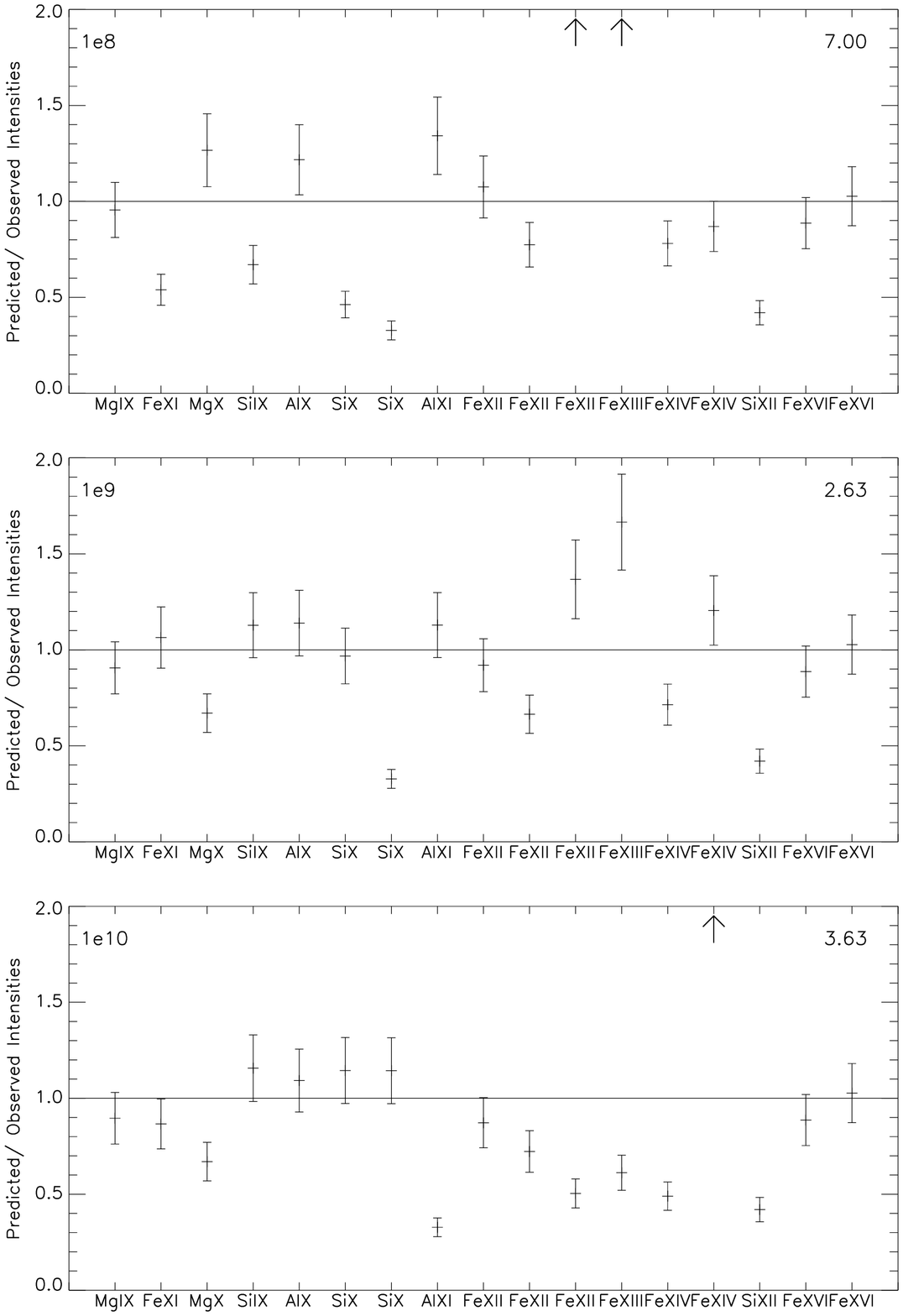}
\caption{Density analysis. Same as Figure 4, but for Loop 3.
}
\end{figure}

\end{document}